\documentclass[twocolumn,aps,prl,footinbib,reprint,longbibliography]{revtex4-1}
\usepackage{comment,xcolor}
\usepackage{amsmath}
\usepackage{mathtools}
\usepackage{amssymb}
\usepackage{amsfonts}
\usepackage{amsthm}
\usepackage{mathrsfs}
\usepackage[all]{xy}
\usepackage[colorlinks=true, citecolor=teal, linkcolor=teal,urlcolor=teal,linktocpage]{hyperref}
\usepackage{tensor}
\usepackage{footnote}
\usepackage{graphicx} 
\usepackage[justification=raggedright,singlelinecheck=false]{caption}
\usepackage{subcaption}
\usepackage{cleveref}
\usepackage{booktabs}
\usepackage{tabularx}

\theoremstyle{plain}
\newtheorem{thm}{Theorem}

\theoremstyle{definition}

\newcommand{\dd}{\mathrm{d}}

\newcommand{\beq}{\begin{equation}}
\newcommand{\eeq}{\end{equation}}

\def\R{\mathcal{R}}
\def\Q{\mathcal{W}}
\def\M{\tilde{M}}

\begin{document}

\title{Local gravitational energy in higher dimensions}
\author{Jinzhao Wang }
\email{jinzwang@phys.ethz.ch}
\affiliation{\small \it Institute for Theoretical Physics, ETH 8093 Z\"urich, Switzerland}

\begin{abstract}
In general relativity, the local gravitational energy is best characterised by the quasilocal mass. The small sphere limit of quasilocal mass provides us the most local notion of gravitational energy. In four dimensions, the limits were shown be the stress tensor in non-vacuum and the Bel-Robinson tensor in vacuum. We study the local gravitational energy in higher dimensions through the lens of the small sphere limits of various quasilocal mass proposals which can be appropriately generalised beyond four dimensions, and report a new quantity $\Q$ which potentially characterises the local gravitational energy content in vacuum. We find that the limits at presence of matter yield the stress tensor as expected, but the vacuum limits are not proportional to the Bel-Robinson superenergy $Q$ in dimensions $n>4$. The result defies the role of the Bel-Robinson superenergy as characterising the gravitational energy in higher dimensions, albeit the fact that it uniquely generalises. More surprisingly, $\Q$ replace the Bel-Robinson superenergy $Q$ in all three quasilocal mass proposals that we study. However, $\Q$ cannot be directly interpreted as a local gravitational energy because of its non-positivity.  The physical meaning of $\Q$ awaits more evidence from investigating other quasilocal masses in higher dimensions. 
\end{abstract}

\maketitle

\emph{Introduction.}---The gravitational field itself carries energy, but it is tricky to locally describe it in general relativity. It is well know that the equivalence principle forbids a covariant stress tensor characterising the energy content of the gravitational field \cite{misner2017gravitation}. Nevertheless, there is no obstruction in giving nonlocal prescriptions and the quasilocal mass (QLM) is such an attempt. Over the past half-century, QLM is an ongoing research subject studied by both physicists and mathematicians \cite{szabados2009quasi,hawking1968gravitational,hayward1994quasilocal,penrose1982quasi,brown1993quasilocal,kijowski1997simple,liu2003positivity,wang2009quasilocal,epp2000angular,bartnik1989new,bousso2019outer}. Nevertheless, QLM is rarely studied in dimensions beyond four. 

Here we make an attempt to investigate the local gravitational energy in higher dimensions, through the lens of quasilocal mass proposals that can be reasonably generalised to higher dimensions. As the gravitational energy density is an invalid notion, the small sphere limit is as local as one can probe about the gravitational energy. It also serves as an important guidance for a sound definition for QLM. Physically, the limit should be proportional to the stress tensor at leading order at the presence of matter or the Bel-Robinson (BR) superenergy $Q_0$ in vacuum \cite{szabados2009quasi}, which we define later. It is natural to expect that the stress tensor will be the leading order characterisation of quasilocal mass or any local gravitational energy in arbitrary dimensions. Furthermore, given that the BR superenergy uniquely generalises to higher dimensions \cite{senovilla2000super}, which we denote as $Q$, one should expect the QLM defined for higher dimensions to reproduce $Q$ as well. In $n=4$, there are many results concerning the small sphere limits of various QLM's. Some notable results concern the Hawking mass by Horowitz and Schmidt \cite{horowitz1982note}, the Brown-York (BY) mass by Brown, Lau and York (BLY) \cite{brown1999canonical}, the Kijowski-Epp-Liu-Yau (KELY) mass by Yu \cite{yu2007limiting} and the Wang-Yau (WY) mass by Chen, Wang and Yau \cite{chen2018evaluatin}. They all exactly agree upon the non-vacuum limit. In vacuum, these QLM's give $\frac1{90}Q_0$ (up to an extra term in cases of KELY and WY) in the small sphere limit via the lightcone cuts.  We generalise these definitions, except for WY mass, to higher dimensions under appropriate assumptions and study their small-sphere behaviours. We find that in non-vacuum there is no surprise that the stress tensor shows up universally. However, in vacuum with $n>4$, a new quantity $\Q$~(\ref{newQ}) instead of the BR superenergy $Q$(\ref{eqn:W}) plays the role of $Q_0$ (\ref{br4d}) in four dimensions. These quantities will be introduced later.

There is a canonical way to evaluate the small sphere limits as proposed by Horowitz and Schmidt \cite{horowitz1982note} in studying the Hawking mass. Let $N_p$ denote the future-directed lightcone located at $p$ generated by null generators $\ell^+$ parameterised by affine parameter $l$. We pick a future-directed timelike unit vector $e_0$ and normalised $\ell^+$ at $p$ by $\langle e_0,\ell^+ \rangle =-1$. The lightcone cut is the family of codimension-two surfaces $S_l$ define as the level sets of $l$ on $N_p$. Hence, $S_l$ implicitly depends on the choice of $(p,e_0)$. The ingoing null generators on $N_p$ are denoted as $\ell^-$ and they are normalised by $\langle \ell^-,\ell^+ \rangle =-1$. The small sphere limit is given by evaluating the QLM on $S_l$ and take $l$ to zero. Note that people use a different small sphere limit in the Riemannian setting \cite{fan2009large,wiygul2018bartnik} and obtain results of the Hawking mass and the BY mass, which are not comparable with results evaluated using the lightcone cuts in the spacetime setting.

It is in the same work of Horowitz and Schmidt \cite{horowitz1982note} that a connection between the quasilocal mass and the BR tensor is discovered. It endows the BR tensor $Q_{abcd}$, more precisely the $Q_{0000}$ component, with the physical meaning of local gravitational superenergy. In the lightcone cut setting, $Q_{0000}$ refers to $Q(e_0,e_0,e_0,e_0)$ and we shall denote it as $Q_0$. It is a `superenergy' but not an energy density due to its different dimension. In fact, using dimensional analysis, one can argue that in four dimensional vacuum any Lorentz invariant quasilocal mass expression for a small sphere must be proportional to $Q_{0}$ at leading order \cite{szabados2009quasi}. This justifies the interpretation of $Q_{0}$ as purely gravitational energy.  It is most conveniently represented in the electromagnetic decomposition ($E,H,D$) of the Weyl tensor. We quickly review the decomposition and the BR superenergy here.

Given some timelike vector $e_0$ at $p$, in adapted coordinates $\{x^0,x^i\}$ where $\partial_{x^0}=e_0$, the Weyl tensor $C$ can be decomposed into spatial tensors
\beq\label{emparts}
E_{ij} := C_{0i0j},\;\;\;\;\; H_{ijk} := C_{0ijk} ,\;\;\;\;\; D_{ijkl} := C_{ijkl},
\eeq
where $E_{ij}$ is the electric-electric part, $H_{ijk}$ is the electric-magnetic part and $D_{ijkl}$ is the magnetic-magnetic part. In four dimensions, the Bel-Robinson tensor~\cite{Bel,robinson1997bel} is
\beq
\begin{aligned}
Q^{(4)}_{abcd} = C_{aecf} C_b{}^e{}_d{}^f + C_{aedf} C_b{}^e{}_c{}^f - \frac{1}{2}g_{ab}C_{cefg}C\indices{_d^{efg}}
\end{aligned}
\eeq
which is defined in a way similar to how the electromagnetic stress tensor is built from the electromagnetic tensor. The BR tensor in four dimensions enjoys many nice properties, such as being traceless, totally symmetric and satisfying a conservation law \cite{senovilla2000super}. Most importantly, it satisfies the dominant property, which means that the tensor $Q^{(4)}_{abcd}$ contracted with any four future directed causal vectors is non-negative. The superenergy is defined as
\beq
Q_0:=Q^{(4)}(e_0,e_0,e_0,e_0)=E^2+B^2, \label{br4d}
\eeq
where $E^2:=E_{ij}E_{ij}, B_{ij}:=\frac{1}{2}\epsilon_{jkl}H\indices{_i_k_l}$. This form suggests the name `superenergy' analogous to the field energy in electrodynamics due to its different dimension. 

Senovilla discovered the following generalisation of the BR tensor in higher dimensions~\cite{senovilla2000super}.
\beq \label{eqn:Tarb}
\begin{aligned}
Q_{abcd} &= C_{aecf} C_b{}^e{}_d{}^f + C_{aedf} C_b{}^e{}_c{}^f 
- \frac{1}{2} g_{ab}  C_{gecf} C^{ge}{}_d{}^f \\
&-\frac{1}{2} g_{cd} C_{aegf} C_b{}^{egf}
+\frac{1}{8} g_{ab} g_{cd} C_{efgh}C^{efgh}.
\end{aligned}
\eeq
and the BR superenergy is defined as 
\beq
Q:=Q(e_0,e_0,e_0,e_0)
\eeq
which is an unique generalisation of the standard BR superenergy $Q_0$ in $n\geq 4$ given the tensor $Q_{abcd}$ is dominant and quadratic in Weyl \cite{senovilla2000super}. Using the Definition \ref{emparts}, one can rewrite $Q$ to be manifestly non-negative
\beq \label{eqn:W}
Q = \frac12\left[E^2+H^2+\frac14 D^2\right],
\eeq
where $E^2:=E_{ij}E_{ij}, H^2:=H_{ijk}H_{ijk}, D^2:=D_{ijkl}D_{ijkl}$. 

Note that $E_{ij}$ is basically the trace of $D_{ijkl}$ and $D_{ijkl}$ contain more information via the off diagonal entries for $n>4$. When $n=4$, they are equivalent and $D^2=4E^2, H^2=2B^2$, so $Q$ equals to $Q_0$. The new vacuum limit we found is given by
\beq \label{newQ}
\Q :=  \frac{(6n^2-20n+8)E^2+6(n-3)H^2-3D^2}{36(n-3)(n-2)(n^2-1)},
\eeq
which only matches with $Q$ when $n=4.$

Even though $Q$ is unique, it may not necessarily acquire the physical meaning of a gravitational energy as $Q_0$ in four dimensions. Our results below show that such a physical characterisation in terms of QLM is indeed missing here, where $\Q$ replaces $Q$.

\emph{Generalisations of QLM.}---We first study a natural $n-$dimensional generalisation of the Hawking mass. Hawking's proposal \cite{hawking1968gravitational} is motivated by the gravitational radiation. For a codimension-2 closed spacelike surface $S$ in a $n$-dimensional spacetime, the Hawking mass is defined as
\beq\label{Hawking}
M_H(S) = \frac{\left(\frac{\text{Vol}(S)}{\Omega_{n-2}}\right)^{\frac{1}{n-2}}}{(n-2)(n-3)\Omega_{n-2}}\int_S \left( \frac{\R}{2} + \frac{n-3}{n-2} \theta^-\theta^+ \right)\dd\sigma
\eeq
where $\sigma$ is the induced volume form on $S,\;\text{Vol}(S)$ is the area of $S=\int_S\dd\sigma$, $\Omega_{n-2}$ is the volume of unit sphere $S^{n-2}$, $\R$ is the Ricci scalar of $S$ and $\theta^\pm:=\nabla_a\ell^{\pm +}$ are the in(out)going null expansions. The same generalisation of the Hawking mass $M_H$ has been studied in \cite{miao2017quasi} and appears in a discussion of quasilocal mass in~\cite{bousso2019outer}. This is a natural generalisation because it retains the properties that the original Hawking mass satisfies (One can find a list of criteria that a sound QLM proposal should comply with in \cite{ChristodoulouYau,szabados2009quasi,wang2015four}). More specifically, $M_H$ reduces to the Misner-Sharp mass for round spheres in spherically symmetric spacetime, vanishes for apparent horizons and has been shown to yield the ADM mass at spacial infinity \cite{miao2017quasi}. By requiring these properties, and our result of the non-vacuum limit in Theorem \ref{thm:hawking} below, the coefficients in $M_H$ are uniquely fixed.

We also study a different class of QLM definitions based upon the Hamilton-Jacobi analysis\cite{brown1993quasilocal,liu2003positivity,wang2009quasilocal,miao2015quasi}. They are the Brown-York mass \cite{brown1993quasilocal} and the Kijowski-Epp-Liu-Yau mass \cite{epp2000angular,kijowski1997simple,liu2003positivity}. One important feature that distinguishes this approach from others is that it requires a flat reference via isometric embedding of the codimension-2 surface to the Minkowski spacetime as the zero-point energy. In particular, we will be considering the lightcone reference, where $S$ is embedded on a lightcone in the Minkowski spacetime. We choose to use such a reference in order for our results to be comparable to earlier works by BLY~\cite{brown1999canonical} and Yu~\cite{yu2007limiting}. In four dimensions, the existence for the lightcone embedding is guaranteed by Brinkmann's result \cite{brinkmann1923riemann}, and more general embedding is guaranteed by Nirenberg \cite{nirenberg1953weyl} and Pogorelov \cite{pogorelov1952regularity}.  In higher dimensions, however, the isometric embedding problem is overdetermined, so generally the reference energy cannot be defined. It is still an open problem to properly generalise the above mentioned Hamilton-Jacobi based proposals to higher dimensions. Nevertheless, for our purposes of looking at the small sphere limit along the lightcone cut specified by $(p,e_0)$, we can still proceed under the assumption that such isometric embeddings do exist for our choices of $(p,e_0)$. Moreover, since only the leading behaviours matter to us, we can ask for the embedding being isometric only approximately up to the perturbative order of interest. The same existence assumptions are held by Miao, Tham and Xie in \cite{miao2017quasi} in investigating the global behaviours of QLM in higher dimensions. 

 The Brown-York mass is not a covariant proposal for QLM and one needs to specify a hypersurface $\Sigma$ where $S_l$ lies on. We choose a particular family of hypersurfaces $\Sigma$, following BLY~\cite{brown1999canonical}, by fixing the normal vector $u$, such that when constrained on a lightcone cut $S_l:=\Sigma \cap N_p,$ it is given by $u := \ell^+/2 +\ell^- $ and the normal of $S_l$ in $\Sigma$ is $v := \ell^+/2-\ell^-.$
The lightcone reference is fixed by BLY requiring the outer expansions $\theta^+:=\nabla_a\ell^{+a}$ being identical at the leading order \cite{brown1999canonical}, and we adopt the same choice for consistency. We henceforth denote all the Minkowski counterparts with a tilde $\tilde{ }$. The embedded $\tilde{S_l}$ sits on a lightcone $\tilde{N}_p$ in Minkowski spacetime $\M^n$ and satisfies: 
\begin{enumerate}
\item[i.] $\tilde{S_l}$ is isometric to $S_l$;
\item[ii.] The outer expansion is the same $\tilde{\theta^+} = \theta^+$.
\end{enumerate}
Once $\tilde{S_l}$ is given, $\tilde{\Sigma}$ is fixed similarly as how $\Sigma$ is fixed from $S_l$ above. The motivation behind BLY's reference choice is that it respects both intrinsically and extrinsically to the physical $S_l$ as much as possible. Alternatively, one can use the Euclidean reference where one embeds $S_l$ to $\mathbf{R}^{n-1}$. However, it is shown by BLY that in four dimensions, the limit deviates from the BR superenergy $Q_0$. We believe it is perhaps a less physical choice as compared to the lightcone reference in this context. 

With the reference settled, we can define the BY mass. Assume the existence of isometric embedding of $S_l$ to a null cone in $\M^n$ satisfying conditions i) and ii), the Brown-York mass for lightcone cuts with respect to the lightcone reference is defined as
\beq
M_{BY}(S_l) := \frac{1}{\Omega_{n-2}(n-2)} \int_{S_l} \tilde{H} - H\; \dd \sigma
\eeq
where $\sigma$ is the induced volume form on $S_l$, $H$ is the mean curvature of $S_l$ in $\Sigma$ and $\tilde{H}$ is the mean curvature of $\tilde{S_l}$ in $\tilde{\Sigma}$. It turns out~\cite{wangtoappear} the formula above with the listed assumptions and choices is a well-defined quasilocal integral, meaning that $\tilde{H}, H$ only depend on the geometric data on $S_l$.

The Kijowski-Epp-Liu-Yau mass is a refinement of the Brown-York proposal in terms of the positivity \cite{liu2003positivity}. Unlike the BY mass, the KELY mass is a covariantly defined QLM, so we do not need to fix any hypersurface $\Sigma$ or $\tilde{\Sigma}$ a priori. In order for the KELY mass to be defined, one needs two conditions~\cite{liu2003positivity} i. the intrinsic Ricci scalar on $S_l$ is positive, $\R(l)>0$; ii. the Mean Curvature vector is spacelike~\footnote{For a codimension-2 surface with normal $n_1,n_2$, the mean curvature vector in this basis has two components given by the trace of extrinsic curvature (mean curvature) associated with $n_1,n_2$, and $K^2$ is independent is the choice of $n_1,n_2$. }, $\langle K,K \rangle >0$. These conditions are guaranteed on lightcone cuts $S_l$ for sufficiently small $l$.  Again, the lightcone reference is chosen as in \cite{yu2007limiting} where $\tilde{S_l}$ is embedded on a lightcone $\tilde{N_p}$ in Minkowski spacetime $\M^n$ and we assume its existence. For lightcone cuts $S_l$, the Kijowski-Epp-Liu-Yau mass is defined as
\beq
M_{KELY}(S_l) := \frac{1}{\Omega_{n-2}(n-2)} \int_{S_l} |\tilde{K}| - |K|\; \dd \sigma
\eeq
where $\sigma$ is the induced volume form on $S$, $K (\tilde{K})$ is the mean curvature vector associated with $S_l (\tilde{S_l})$, and $|\cdot |=\sqrt{\langle\cdot,\cdot \rangle}$ denotes the metric norm.

\emph{Results.}---Our main results are:
\begin{thm}\label{thm:hawking}
Let $S_l$ be the family of surfaces shrinking towards $p$ along lightcone cuts defined with respect to $(p,e_0)$ in a $n$-dimensional spacetime, the limits of the Hawking mass as $l$ goes to $0$ are
\begin{enumerate}
\item In non-vacuum,
\beq
\lim_{l\rightarrow 0}l^{-(n-1)}M_{H} = \frac{\Omega_{n-2}}{n-1}T(e_0,e_0).\label{hawkingnonvac0}
\eeq
\item In vacuum or the stress tensor $T$ vanishes in an open set containing $p$,
\beq
\lim_{l\rightarrow 0}l^{-(n+1)}M_{H} = \Q. \label{hawkingvac0}
\eeq
with $\Q := \frac{(6n^2-20n+8)E^2+6(n-3)H^2-3D^2}{36(n-3)(n-2)(n^2-1)}.$
\end{enumerate}
where the tensors $T,\Q,E,H,D$ are evaluated at $p$.
\end{thm}

Note it turns out that the leading order is $l^{n\mp 1}$ so we regularize $M_H$ by multiplying with the reciprocal in order to pull out the leading factor.  As a corollary, one obtains the $n=4$ result by Horowitz and Schmidt \cite{horowitz1982note}. We see that the non-vacuum case (\ref{hawkingnonvac0}) agree with our expectation. The coefficient in front of $T(e_0,e_0)$ is exactly the volume of a flat unit sphere enclosed by $S_l$, and thus together it characterises the dominant matter energy content within the small sphere. However, the vacuum limit (\ref{hawkingvac0}) is not proportional to the BR superenergy $Q$ in any dimensions $n>4$. Note that the area of the lightcone cut $S_l$ itself has a deficit due to curvature that is proportional to $Q_0$ in four dimensions \cite{brown1999canonical}, but it is not proportional to $Q$ in higher dimensions \cite{wang2019geometry}. Here, we see the behaviour of the QLM is in line with the area deficit. Rather surprisingly, the exactly same limit is obtained for the BY mass in higher dimensions.
\begin{thm}\label{thm:BY}
Let $S_l$ be the family of surfaces shrinking towards $p$ along lightcone cuts defined with respect to $(p,e_0)$ in a $n$-dimensional spacetime, and assume the isometric embedding into the lightcone reference exists, the limits of the Brown-York mass as $l$ goes to $0$ are
\begin{enumerate}
\item In non-vacuum,
\beq
\lim_{l\rightarrow 0}l^{-(n-1)}M_{BY} = \frac{\Omega_{n-2}}{n-1}T(e_0,e_0).\label{bynonvac}
\eeq
\item In vacuum or the stress tensor $T$ vanishes in an open set containing $p$,
\beq
\lim_{l\rightarrow 0}l^{-(n+1)}M_{BY} = \Q .\label{byvac}
\eeq
\end{enumerate}
\end{thm}

As noted by BLY \cite{brown1999canonical}, it is rather surprising that in four dimensions, both the Hawking mass and the BY mass yield the same vacuum limit as the two proposals are constructed from two totally different approaches. Here we choose the same lightcone reference as BLY, and we see that their small sphere limit agree in all dimensions. Our results suggest that the generalised Bel-Robinson superenergy $Q$, though unique, does not retain its gravitational energy interpretation beyond four dimensional spacetime. Nevertheless, we are reluctant to refer the new vacuum limit $\Q$ as a new candidate for the gravitational superenergy because this quantity is not always positive. In particular, whenever the magnetic-magnetic component of the Weyl tensor $D$ dominates over the other two contributions under some choice of $(p,e_0)$, we have a negative vacuum limit. This confirms the fact that the Hawking mass and the BY mass are plagued with non-positivity issues. In higher dimensions, it is more serious that non-positivity is manifested even in small lightcone cuts for some choices of $(p,e_0)$, which is not the case for $n=4$.

\begin{thm}\label{thm:LY}
Let $S_l$ be the family of surfaces shrinking towards $p$ along lightcone cuts defined with respect to $(p,e_0)$ in a $n$-dimensional spacetime, and assume the isometric embedding into the lightcone reference exists, the limits of the Kijowski-Epp-Liu-Yau mass as $l$ goes to $0$ are
\begin{enumerate}
\item In non-vacuum,
\beq
\lim_{l\rightarrow 0}l^{-(n-1)}M_{KELY} = \frac{\Omega_{n-2}}{n-1}T(e_0,e_0).\label{hawkingnonvac}
\eeq
\item In vacuum or the stress tensor $T$ vanishes in an open set containing $p$,
\beq
\lim_{l\rightarrow 0}l^{-(n+1)}M_{KELY} = \Q-\frac{(n^2-n-3)E^2}{12(n^2-1)(n-2)^2(n-3)^2}.\label{hawkingvac}
\eeq
\end{enumerate}
\end{thm}

In four dimensions, the vacuum limit is the BR superenergy $Q_0$ with an extra term proportional to $E^2$ \cite{yu2007limiting}. We see that in higher dimensions, the same pattern holds and $Q_0$ generalises to $\Q$ instead of $Q$ as for the Hawking mass and the BY mass. When $n=4$, the vacuum limit is positive but fails to remain positive in higher dimensions if $D^2$ dominates. 

The detailed proof of the claimed theorems will be provided in a separate paper~\cite{wangtoappear}. We briefly outline the strategy here. By fixing appropriate gauges for the null generators, we explicitly work out the perturbative geometry of the lightcone cuts in a curved $n$-dimensional spacetime. They are the induced metric, intrinsic Ricci scalar, extrinsic curvature, mean curvature, expansion and shear assoicated with $S_l$. These allow us to compute the quasilocal integrals in a perturbative expansion. 

\emph{Discussion.}---We discovered a new quantity $\Q$ represented in terms of the electromagnetic decompositions of the Weyl tensor. It replaces the BR superenergy $Q$ in the context of QLM. Albeit that $\Q$ is not positive-definite, we believe that $\Q$ nevertheless characterises the local gravitational energy content, perhaps in an elusive way. Further investigation is certainly needed to clarify its physical meaning. One can try to study other QLM proposals. The Wang-Yau mass \cite{wang2009quasilocal} is the most recent QLM proposal using the Hamilton-Jacobi approach. It overcomes all the shortcomings of the BY and the LY proposals \cite{miao2015quasi}. However, its potential generalisation to higher dimensions is obscure as the definition relies on the data of isometric embedding so one cannot simply assume its existence, even just for the small sphere calculations. It would be insightful to see how the Wang-Yau proposal can be extended to higher dimensions and whether $\Q$ appears in the limit. There is another QLM that can be naturally generalised to higher dimensions, the Bartnik mass \cite{bartnik1989new}. Roughly speaking, it is defined as the infimum among the ADM masses evaluated on all horizon-free spacetime extensions of a quasilocal region. Its positivity and rigidity follow from the positive mass theorem, and therefore it is a promising candidate of QLM alongside the WY mass. However, the Bartnik mass is difficult to evaluate generally, especially in the spacetime case. The small sphere limit of the Bray's modification, Bartnik-Bray mass, has been evaluated in a time-symmetric (Riemannian) setting \cite{wiygul2018bartnik}, but they are not comparable with results in the spacetime setting. It would be very interesting to know how to compute its small sphere limit along lightcone cuts and compare to our results here. 

Meanwhile, one perhaps needs more input from high energy physics and quantum gravity to have a comprehensive understanding of the local gravitational energy. Assuming the quantum mechanical nature of spacetime, the QLM in vacuum should account for the quantum excitations of genuine gravitational degrees of freedom. One such probe is the holographic duality. Recently, there is a holographic QLM proposal defined using the outer entropy \cite{bousso2019outer}, a coarse-grained entropy associated with an extremal surface introduced by Englehardt and Wall \cite{engelhardt2018decoding, engelhardt2019coarse}. The holographic QLM is interesting as it is defined for arbitrary dimensions and the holography setup suggests a quantum interpretation of the QLM, a purely classical bulk construction, in terms of the entanglement entropy defined on the boundary. However, it has not been checked if the proposal satisfies all the criteria to be qualified as a QLM, such as its large and small sphere limits, and we shall leave the investigations about the holographic QLM to future works.

\emph{Acknowledgement}---We thank Jos\'e M.M.Senovilla for clarifying the Bel-Robinson superenergy and Renato Renner for useful comments. This work is supported by the Swiss National Science Foundation via
the National Center for Competence in Research “QSIT”, and by the Air Force Office of Scientific Research (AFOSR) via grant FA9550-16-1-0245. 

\bibliography{HM2}

\end{document}